\documentclass[12pt]{iopart}
\usepackage{iopams}  
\usepackage{graphicx}
\usepackage{hyperref}

\begin{document}

\title[Collective oscillations of excitable elements]{Collective oscillations of
  excitable elements: order parameters, bistability and the role of
  stochasticity}

\author{Fernando Rozenblit and Mauro Copelli}

\address{Departamento de F{\'\i}sica, Universidade Federal de
  Pernambuco, 50670-901, Recife-PE, Brazil}
\eads{\mailto{rozenblit@df.ufpe.br}, \mailto{mcopelli@df.ufpe.br}}

\begin{abstract}
  We study the effects of a probabilistic refractory period in the
  collective behavior of coupled discrete-time excitable cells
  (SIRS-like cellular automata). Using mean-field analysis and
  simulations, we show that a synchronized phase with stable
  collective oscillations exists even with non-deterministic
  refractory periods. Moreover, further increasing the coupling
  strength leads to a reentrant transition, where the synchronized
  phase loses stability. In an intermediate regime, we also observe
  bistability (and consequently hysteresis) between a synchronized
  phase and an active but incoherent phase without oscillations. The
  onset of the oscillations appears in the mean-field equations as a
  Neimark-Sacker bifurcation, the nature of which (i.e. super- or
  subcritical) is determined by the first Lyapunov coefficient. This
  allows us to determine the borders of the oscillating and of the
  bistable regions. The mean-field prediction thus obtained agrees
  quantitatively with simulations of complete graphs and, for random
  graphs, qualitatively predicts the overall structure of the phase
  diagram. The latter can be obtained from simulations by defining an
  order parameter $q$ suited for detecting collective oscillations of
  excitable elements. We briefly review other commonly used order
  parameters and show (via data collapse) that $q$ satisfies the
  expected finite size scaling relations.
\end{abstract}

%\noindent{\it Keywords}: synchronization, excitable elements,
%nonequilibrium phase transitions, random graphs, nonlinear dynamics,
%hysteresis, collective oscillations, SIRS, epidemiology, neuroscience
%\submitto{J. Stat. Mech.}
\maketitle

\tableofcontents

\section{Introduction}

Understanding collective oscillations of coupled nonlinear elements
remains a challenge from both theoretical and experimental
viewpoints. From the theoretical side, much progress has been
accomplished since the seminal works of Winfree and
Kuramoto~\cite{Winfree67,Kuramoto84,Strogatz00,Pikovsky01,Acebron05}
on coupled oscillators, upon which recent literature has expanded to
include effects of e.g. complex topologies~\cite{Arenas08} and
noise~\cite{Risler04,Risler05,Wood06a,Wood06b,Wood07a,Wood07b,Khoshbakht08}. From
the experimental side, the subject has a longstanding importance in
neuroscience: collective neuronal oscillations stood for a long time
as candidates for a solution of the so-called binding
problem~\cite{Singer99}, but emphasis has recently shifted to
attentional processes~\cite{Uhlhaas09}.

Here we are interested in collective oscillations of units which are
excitable, i.e. not intrinsically oscillatory. This topic has been
experimentally observed in a variety of scenarios, from
neuroscience~\cite{Shew09,Ko2010} to chemistry~\cite{Taylor09}, but
theoretical approaches have been relatively
scarce~\cite{Kuperman01,Girvan02,Gade05,Kinouchi06a,Assis09,Liu09,McGraw10}.
In particular, it is not entirely clear to which extent these
oscillations are robust with respect to noise.  On the one hand,
recent studies have shown sufficient conditions for the onset of
global oscillations of {\em deterministic\/} excitable units with {\em
  noisy coupling\/}, emphasizing e.g. the interplay between coupling
strength and characteristic time scales of the
units~\cite{Girvan02,Gade05}, or the importance of the topology of the
network~\cite{Kuperman01}. On the other hand, the
susceptible-infected-recovered-susceptible (SIRS) model on a lattice
(i.e. a minimum three-state excitable model) has been thoroughly
studied, with several results strongly suggesting that its {\em
  stochastic\/} Markovian version does not yield sustained
oscillations~\cite{Rozhnova09,Rozhnova09b,Rozhnova09c,Assis09} (for
non-Markovian models, see e.g.~\cite{Lindner04,Abramson10a}).

This raises the question whether it is possible to find sustained
global oscillations in a network of excitable units whose intrinsic
dynamics (not only the coupling) is non-deterministic. We therefore
propose and study a simple probabilistic model which has a
well-defined deterministic limit. To study the phase transitions in
the model, we employ an order parameter specifically tailored to
assess collective oscillations of excitable systems. These are
described in section~\ref{model}. We study complete graphs as well as
random graphs, comparing mean-field calculations with
simulations. Results are presented in
sections~\ref{results.cg}~and~\ref{results.rg}, while
section~\ref{conclusions} brings our concluding remarks.

\section{\label{model}Model}

\subsection{Excitable cellular automata}

\begin{figure}[ht!]
  \centering
  \includegraphics[width=0.3\columnwidth]{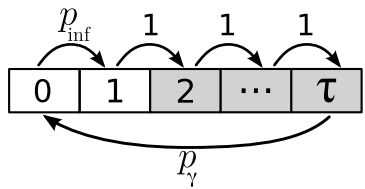}
  \caption{\label{fig:model}Single-cell dynamics. $p_{inf}$ is the
    probability of activation ($0 \rightarrow 1$) from neighbours,
    $p_{\gamma{}}$ is the probability of transitioning from the relative
    refractory state~($\tau$) to the rest state~($0$). Light
    gray states are refractory. All other transitions are deterministic.}
\end{figure}

The minimum model of an excitable system consists of three states,
representing quiescence (state 0), excitation (state 1) and
refractoriness (state 2)~\cite{Lindner04} (a prototypical example
being the SIRS model). As shown by Girvan et al., however, such a
cyclic three-state deterministic cellular automaton fails to exhibit
sustained collective oscillations. For them to become stable, their
model needs at least 2 refractory states~\cite{Girvan02}. To obtain an
arbitrary number of refractory states, for each site $j = 1, ..., N$,
let $s_j = 0, 1, 2, ..., \tau$ be the consecutive states of the unit
(out of the $\tau + 1$ states, the last $\tau-1$ are
refractory~\cite{Girvan02}, see~\fref{fig:model}).

The cellular automaton version of the probabilistic SIRS model (also
called the probabilistic Greenberg-Hastings model~\cite{Greenberg78})
corresponds to $\tau=2$, with intrinsic transitions $1\rightarrow 2$
and $2\rightarrow 0$ governed by constant probabilities, whereas the
transition $0\rightarrow 1$ occurs with a probability that usually
increases linearly with the number of excited neighbors~\cite{Assis08}
(for a study with nonlinear coupling, see~\cite{Assis09}). In the
model studied by Girvan et al., on the other hand, all intrinsic
transitions ($s_j \rightarrow (s_{j}+1) \mbox{ mod } (\tau+1)$, $s_j\neq
0$) are deterministic.

Here we study an intermediate variant of these models, where all
intrinsic transitions are deterministic but the last one, which occurs
with probability $p_\gamma$ (see~\fref{fig:model}). The idea is to
have a minimum model (lest the number of additional parameters becomes
too large) which incorporates non-determinism in the intrinsic
dynamics. The choice to make the transition from the last refractory
state probabilistic is natural and comes from neuroscience: neuronal
dynamics depend on ionic channels which are
stochastic~\cite{Koch,Carelli05}, so that a neuron may or may not fire
when stimulated at the {\em end\/} of its refractory period (the
so-called relative refractory period)~\cite{Koch}.

\subsection{\label{coupling}Coupling}

The only transition that still needs to be described is the excitation
process $0\rightarrow 1$. We assume each site $j$ is symmetrically
connected with $k_j$ other sites. Each active site has a probability
$\sigma/K$ of activating a resting neighbour, where $\sigma$
is a control parameter (which corresponds to the system branching
ratio~\cite{Harris,Kinouchi06a}) and $K$ is the average connectivity
($K = \left<k_j\right>$). 
The fraction of active sites 
\begin{equation}
  \label{eq:Pt1}
P_t(1) \equiv \frac{1}{N}\sum_{j=1}^N \delta_{1,s_j(t)}  
\end{equation}
is used to measure network activity at time $t$. At $\sigma = 1$ the
model shows a transition from an absorbing to an active state, but
without sustained oscillations~\cite{Kinouchi06a}. With deterministic
units ($p_\gamma = 1$), the system undergoes a transition to the
oscillatory regime at $\sigma = \sigma_c(p_\gamma = 1) > 1 $, which
persists indefinitely if $\sigma$ is further
increased~\cite{Girvan02,Kinouchi06a}.

\begin{figure}[ht!]
  \centering
  \includegraphics[width=0.8\columnwidth]{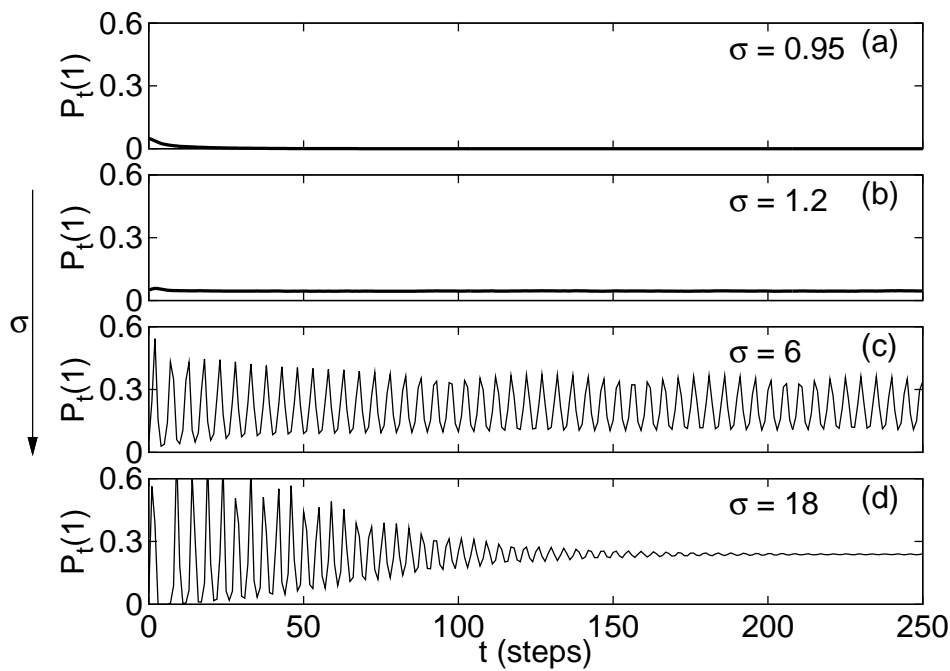}
  \caption{\label{fig:variousSigma}Time series for $p_\gamma = 0.85$
    and increasing values of $\sigma$ on a complete graph with $N =
    5\times{}10^5$ sites. Respectively for increasing $\sigma$
    (top-bottom): (a) absorbing (not-active) state, (b) active state without
    oscillations, (c) with oscillations and (d) again active without
    oscillations. $P_0(0) = 0.95$, $P_0(1) = 0.05$. $\tau = 3$.}
\end{figure}

To motivate the analysis to be developed in section~\ref{results.cg},
\fref{fig:variousSigma} shows examples of single-run results in a
complete graph with $N= 5\times 10^5$ for $p_\gamma= 0.85$ and
increasing values of $\sigma$. For probabilistic units, we observe the
transition to an active but nonoscillating state at $\sigma=1$
[\Fref{fig:variousSigma}(a)-(b)] and the second transition to an
oscillating state for larger $\sigma$
[\Fref{fig:variousSigma}(b)-(c)]. Contrary to what is observed in the
deterministic model, however, in the probabilistic model this second
transition is {\em reentrant\/} with respect to the coupling strength
$\sigma$, as shown in \fref{fig:variousSigma}(c)-(d).

The nature of this reentrant transition will be clarified in
section~\ref{results.cg}. In order to analyze it properly, though, one
needs to define an adequate order parameter to detect synchronization
among excitable elements.

\subsection{\label{orderparameter}Order parameters}

Most studies of synchronization employ the Kuramoto order
parameter~\cite{Kuramoto84,Strogatz00}, which corresponds to the time
and ensemble average of the norm of the complex vector
\begin{equation}
  \label{eq:complexKuramoto}
  Z(t)  \equiv \frac{1}{N} \sum^{N}_{j = 1}{e^{i\theta{}_j(t)}} \; ,
\end{equation}
where $\theta{}_j = 2\pi{}s_j/(\tau{}+1)$. Note that $Z$ corresponds
to the center of mass of the phases of the units. This works fine when
the system is composed of coupled {\em uniform\/} oscillators, because
rotational symmetry will ensure that, in the absence of sustained
collective oscillations, the order parameter vanishes in the
thermodynamic limit. Consider, however, the present case of excitable
elements. The trivial absorbing state ($s_j=0$, $\forall j$), which is
always a collective solution of the dynamics, yields a nonzero (in
fact, maximum!) Kuramoto order parameter. Indeed, in the absorbing
state units are all ``perfectly synchronized'' in the sense that they
always have the same state. But this is clearly not what one wants to
detect. This problem persists for $\sigma \gtrsim 1$ (below the onset
of collective oscillations), where a small fraction of the units are
active (on average), whereas most remain quiescent. In that case, $Z$
has a {\em constant\/} bias towards the absorbing state, around which
it will fluctuate.

Collective oscillations correspond to rotations of $Z$ which may be
misdetected in the averaging procedure owing to a lurking constant
vector. One possibility which has been used to avoid the weight of the
absorbing state is excluding terms with $s_j=0$ from the sum in
eq.~\eref{eq:complexKuramoto}~\cite{Kuperman01,Gade05}. Another
strategy makes use of the standard deviation (measured along time) of
$P_t(1)$ to detect oscillations~\cite{GarciaOjalvo2004}, though
neither procedure is easily extensible to continuous-phase systems.

To account for a system of continuous-phase units which may have an
arbitrary number of preferred phases, one could employ the angular
momentum $L \equiv X\partial_t Y - Y\partial_t X$, where
$Z=X+iY$~\cite{Kuramoto92a,Ohta08}. In our cellular automata, this
would require the discretization of the time derivative, which could
introduce unnecessary numerical errors. Alternatively, Shinomoto and
Kuramoto have previously proposed
\begin{equation}
  \label{eq:parametroideal}
  \tilde{q} \equiv \langle \left| Z - \langle Z \rangle_t
  \right|\rangle_t \; ,
\end{equation}
which amounts to subtracting the constant bias from $Z$ before the
averaging procedure~\cite{ShinomotoKuramoto86}. However, this can be
computationally expensive, requiring the storage of the whole time
series. Making use of a similar idea, but with much less computational
bookkeeping, in the following we will characterize collective
oscillations via the order parameter
\begin{equation}
  \label{eq:syncACR}
  q = \sqrt{\langle\left| Z - \langle Z \rangle_t \right|^2\rangle_t}
  = \sqrt{\langle\left|Z\right|^2\rangle_t - \left|\langle Z \rangle_t \right|^2} \; ,
\end{equation}
which can be seen as a generalized standard deviation of
$Z(t)$. Differently from $\tilde{q}$, obtaining $q$ is computationally
inexpensive, as the means over time are now separated and may be
calculated along with the simulation. In section~\ref{fsizescaling} we
will show that $q$ satisfies scaling relations near a phase
transition, as expected for an order parameter.

One may grasp intuition about $q$ by considering the different time
series in~\fref{fig:variousSigma}. Let $P_1^*$ be the stationary value
$\lim_{t\to\infty}\langle P_t(1)\rangle_t$, which is an order
parameter in its own right, measuring whether or not the network is
active~\cite{Kinouchi06a}. In figs.~\ref{fig:variousSigma}(a) and (b),
although the system goes from $P_1^*=0$ to $P_1^*\neq 0$ as the system
goes from an absorbing to an active fixed point, both have $q=0$,
because there are no oscillations after the
transient. In~\fref{fig:variousSigma}(c), on the other hand, we have
both $P_1^*\neq 0$ and $q>0$, as the oscillatory state becomes stable
after increasing $\sigma$. Finally, for~\fref{fig:variousSigma}(d),
the oscillations are unstable, $P_1^*\neq 0$ and $q=0$.

\section{\label{results.cg}Complete graph}

We start with the complete graph because it is presumably the topology most
prone to exhibiting stable collective oscillations. Besides, it allows
a comparison between simulations and analytical results (see
below). From now on, we will focus on the effect of the probabilistic
dynamics ($p_\gamma$) on synchronization and will fix $\tau=3$. 

\subsection{\label{results.cg.sim}Simulations}

We have simulated complete graphs ($k_j = K = N-1$) with sizes varying
from $N=10^5$ to $N=10^6$. At each time step, intrinsic transitions
occur as described in section~\ref{model}. The transition $0\to 1$ is
governed by the number $N_1(t)$ of active ($s_i = 1$) sites at time
$t$, each of which can activate a quiescent cell with probability
$\sigma/N$. Therefore, the probability of a quiescent cell being
activated by at least one of its $N_1(t)$ active neighbours is
\begin{equation}
  p(0\to 1) = P_{inf}(t) = 1 - \left( 1 - \frac{\sigma}{N} \right)^{N_1(t)}\;,
\end{equation}
which renders the simulations relatively simple despite the large
system sizes. Finally, initial conditions must be chosen to avoid
large amplitudes during the transient, which could throw the system
into the absorbing state~\cite{Girvan02}. Apart from that, the effects
we report below are robust with respect to the initial conditions and
we have arbitrarily fixed $P_0(0)=0.8$ and $P_0(1)=0.2$.

\begin{figure}[ht!]
\begin{center}
    \includegraphics[width=0.65\columnwidth]{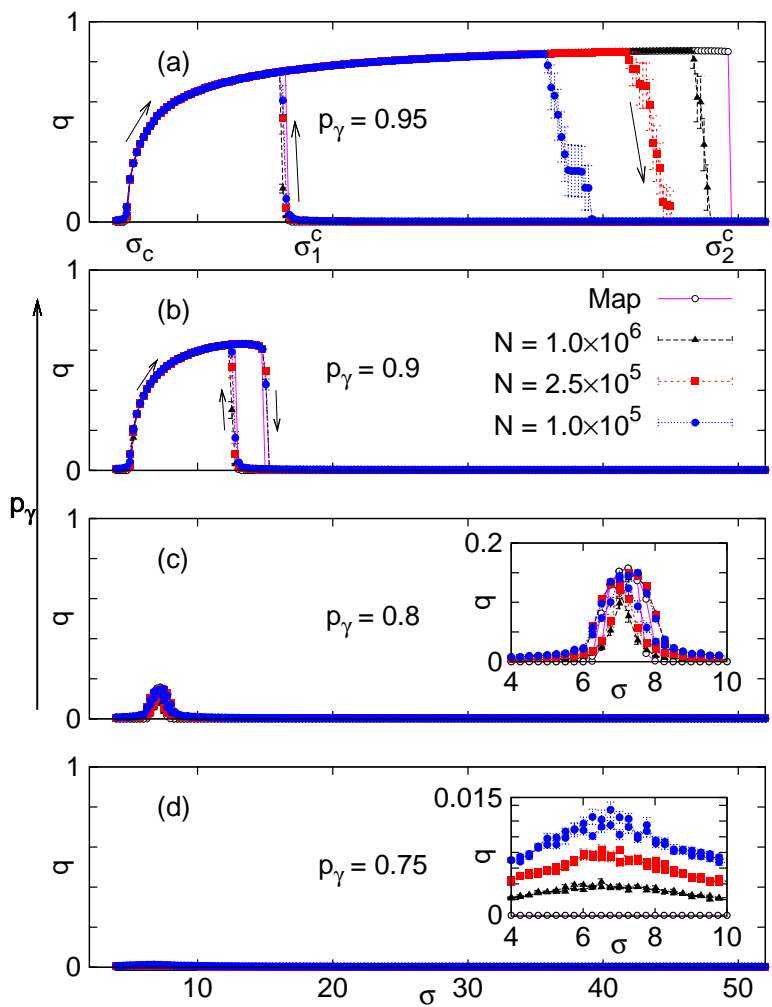}
    \caption{\label{fig:histereses} Hysteresis loop for the mean-field
      solution (to be presented in section~\ref{MF}) and complete
      graph ($N = 10^5$, $2.5\times{}10^5$ and $10^6$) with
      $p_{\gamma} = 0.95$, $0.9$, $0.8$ and $0.75$. $t_{max} =
      1\times{}10^3$ steps ($t_{trans}=500$) and
      $\delta\sigma=0.05$. Mean (symbols) and standard errors (bars)
      calculated over $10$ runs. Insets zoom into the interesting
      region $4 \leq \sigma\leq 10$. Note that fluctuations in the
      inset of (d) decrease with increasing system size, where the
      mean-field approximation (section~\ref{MF}) predicts $q=0$ in
      the thermodynamic limit (see also section~\ref{fsizescaling}).}
\end{center}
\end{figure}

To illustrate the kind of reentrant transition exemplified in
\fref{fig:variousSigma}, we show in \fref{fig:histereses}(a) the order
parameter $q$ as a function of the coupling parameter
$\sigma$. Starting at some $\sigma_{min}>1$, for each value of
$\sigma$ we let the system evolve during a transient of $t_{trans}$
time steps, after which we start measuring the order parameter $q$ up
to $t=t_{max}$ time steps. We then increase $\sigma$ by a constant
amount $\delta\sigma$ and repeat the procedure, with the initial
condition of the system for each value corresponding to the final
condition of the preceding value. Both constants are chosen so the
rate of change is small ($\delta\sigma{}/t_{max} \ll 1$). After a
maximum value $\sigma_{max}$ is reached, $\sigma$ is sequentially
decreased by the same amount $\delta\sigma$ down to $\sigma_{min}$.

As shown in \fref{fig:histereses}(a), the reentrance of the transition
to collective oscillations is captured by the non-monotonic
behavior of $q(\sigma)$, which departs from zero at some lower value
$\sigma_c(p_\gamma)$ and returns to zero at some upper value
$\sigma_2^c(p_\gamma)$. Moreover, we have found that while the first
transition is always continuous, the second transition can be
discontinuous. The fingerprint of the discontinuity is the hysteresis
observed in the order parameter: above $\sigma_2^c$, the only stable
state of the system has constant (but nonzero) $P_t(1)$, thus no
oscillations ($q=0$). If we decrease $\sigma$, oscillations do not
reappear at $\sigma_2^c$, but rather at a lower value
$\sigma_1^c$. There is therefore a region of bistability $\sigma\in
[\sigma_1^c,\sigma_2^c]$ in parameter space where collective
oscillations ($q>0$) can coexist with an active ($P_t(1) > 0$) but
non-oscillating ($q=0$) state. As it turns out [see
\Fref{fig:histereses}(a)], for $p_\gamma=0.95$ the size of this
bistable region is rather sensitive to the system size. Smaller
systems tend to be perturbed away from collective oscillations by
larger fluctuations, leading to smaller hysteresis cycles. 

As $p_\gamma$ is decreased, the mean and variance of the refractory
periods of the units increase, rendering the whole system
noisier. This might be the explanation for the result in
\fref{fig:histereses}(b), which shows a smaller reentrant region with
oscillations ($q>0$). The width of the hysteresis cycle also
decreases, with both $\sigma_1^c$ and $\sigma_2^c$
decreasing. Furthermore, $\sigma_2^c$ (along with the width of the
hysteresis cycle) becomes less sensitive to the system size, which
could be due to the variance of the refractory periods overcoming the
effects of small-size fluctuations. Albeit subtly, $\sigma_c$ also
increases slowly with decreasing $p_\gamma$, as will be seen
in~\fref{fig:diagrama}.

Further decreasing $p_\gamma$ [\Fref{fig:histereses}(c)], the bistable
region vanishes, whereas the transition to a collectively oscillating
state remains. Finally, for sufficiently small $p_\gamma$, collective
oscillations are no longer stable [\Fref{fig:histereses}(d)]. 

In the following, we will show that these transitions can be
quantitatively reproduced by a low-dimensional mean-field
analysis. For a controlled comparison, $\sigma_c$ and $\sigma_1^c$
were heuristically defined as the values of $\sigma$ (averaged over
$n$ runs) where $q$ first rose above some threshold value $q_{min}
\propto 1/\sqrt{N}$, respectively for increasing and decreasing values
of $\sigma$. On the other hand, $\sigma_2^c$ was defined as the value
of (increasing) $\sigma$ where $q$ first fell below $q_{min}$.

\subsection{\label{MF}Mean-field analysis}

In the following we apply the standard mean-field (MF)
approximation~\cite{Marro99} to the equations governing our system. We
follow closely the steps of
Refs.~\cite{Girvan02,Furtado06,Kinouchi06a}, where every site is
considered to have $K$ neighbors, a fraction $P_t(1)$ of which is
excited at time $t$. If a given site is at rest [with probability
$P_t(0)$], the probability of it becoming excited by at least one of
its excited neighbors is
\begin{equation}
\label{eq:pinf}
P_{inf}(t) =  1 - \left( 1 - \frac{\sigma  P_t(1)}{K} \right)^{K}\;.
\end{equation}
The dynamics of the system is then described by the following closed
set of equations:
\begin{eqnarray}
\label{eq:MFmap}
P_{t+1}(0) &=& p_{\gamma}P_t(\tau) + \left(1-P_{inf}(t)\right) P_t(0)\label{eq:MFredundant} \\
P_{t+1}(1) &=& P_{inf}(t) P_t(0) \\
%P_{t+1}(2) &=& P_t(1)\\ 
%\ & \vdots &\nonumber \\
P_{t+1}(s) &=& P_t(s - 1) \ \ \ \ \ \ \ \ \ \ \ \ \ \ \ \ \ \ \ \ (2 \leq s \leq \tau-1)\label{eq:MFDeterm}\\ 
%\ & \vdots &\nonumber \\
P_{t+1}(\tau) &=& P_t(\tau - 1) + (1 - p_{\gamma})P_t(\tau)\; , \label{eq:MFProbab}
\end{eqnarray}
where the normalization condition
\begin{equation}
\label{eq:norm_cond}
P_t(0) = 1 - \sum_{s = 1}^{\tau}{P_t(s)}\\
\end{equation}
renders \eref{eq:MFredundant} redundant and reduces the system to a
$\tau$-dimensional map~\cite{Furtado06}. Therefore, increasing the
duration of the refractory period amounts to an increase in the
complexity of the mean-field calculations.

For the complete graph, $K = N-1$ and mean field is exact. In the
thermodynamic limit, \eref{eq:pinf} becomes
\begin{equation}
\label{eq:pinf_limit}
\lim_{N \to \infty} P_{inf} = 1 - e^{-\sigma{}P_t(1)} \; ,
\end{equation}
which, from \eref{eq:MFmap} and \eref{eq:norm_cond}, leads to
\begin{equation}
P_{t+1}(1) = (1 - e^{-\sigma{}P_t(1)})\left[1 - \sum_{s =
    1}^{\tau}{P_t(s)}\right]\; .
\label{eq:MFActive} \\
%P_{t+1}(s) &=& P_t(s - 1) \ \ \ \ \ \ \ \ \ \ \ \ \ \ \ \ \ \ \ \ (2 \leq s \leq \tau-1)\label{eq:MFDeterm}\\ 
%P_{t+1}(\tau) &=& P_t(\tau - 1) + (1 - p_{\gamma})P_t(\tau)
\end{equation}
In other words, in the mean field approach the state of the system is
completely described by $\vec{P}_t\equiv \left(
  P_t(1),P_t(2),\ldots,P_t(\tau)\right)^T$, which evolves according to
$\vec{P}_{t+1} = \vec{F}(\vec{P}_{t})$. Note that $F_1$ is the only
component of $\vec{F}$ which is nonlinear [see~\eref{eq:MFActive}].

From~\eref{eq:MFDeterm}, the fixed point for any $2 \leq s \leq
\tau-1$ is clearly $P_s^* \equiv P_{\infty}(s) =
P_{\infty}(s-1)=\cdots=P_1^*$ which, upon substitution
in~\eref{eq:MFProbab}, gives $P_{\tau}^* = P_1^*/p_{\gamma}$. Finally,
in its steady state, \eref{eq:MFActive}~becomes:
\begin{eqnarray}
P_1^* &=& (1 - e^{-\sigma{}P_1^*})\left[1 - \left(\tau{}-1 + \frac{1}{p_{\gamma}}\right)P_1^*\right] \;,
\end{eqnarray}
which can be numerically solved. Expanding near $P_1^*\simeq 0$ (note
that $P_1^*=0$ is always a solution), one easily obtains the
transition from an absorbing to an active (steady) state at
$\sigma=1$~\cite{Kinouchi06a}. The collective oscillations appear when
the active state becomes unstable.

\subsection{Linear stability}

Considering a small perturbation $\eta_t(s)$ such that $P_t(s) = P_s^* +
\eta_t(s)$, the linearized dynamics can be written as $\vec{\eta}_{t+1} =
A\vec{\eta}_{t}$, where $A_{ij} = \left. \partial F_i / \partial P_t(j)
\right|_{\vec{P}^*}$ is the Jacobian matrix calculated at the fixed point:

\begin{equation}
\label{eq:jacobian}
A = \left(\begin{array}{ccccc}
g(\sigma{},P_1^*) & (e^{-\sigma{}P_1^*} - 1) &  \cdots  & (e^{-\sigma{}P_1^*} - 1) & (e^{-\sigma{}P_1^*} - 1) \\
1 & 0 &  \cdots & 0 & 0 \\
0 & 1 &  \cdots & 0 & 0 \\
&  & \ddots & & \\
0 & 0 & \cdots  &  1 & (1 - p_\gamma) \end{array}\right)\;,
\end{equation}
and $g(\sigma{},P_1^*) = e^{-\sigma{}P_1^*} - 1 +
\sigma{}e^{-\sigma{}P_1^*}[1 - (\tau{} - 1 + 1/p_{\gamma})P_1^*]$. The
eigenvalues $\{\mu_j\}_{j=1}^\tau$ of $A$ determine whether
$\vec{P}^*$ is stable ($\max_{j} |\mu_j|<1$) or unstable ($\max_{j}
|\mu_j|>1$) (for simplicity, in the following we employ $\mu \equiv \mu_k$, where
$k=\mbox{argmax}_j |\mu_j|$). 

We expect to pinpoint the transition to the oscillatory state by
looking for a Neimark-Sacker~(NS) bifurcation in the mean field
equations (which is the discrete-time analog of the Andronov-Hopf (AH)
bifurcation in continuous time~\cite{Kuznetsov98}). In other words, for
fixed $p_\gamma$, we have $\left|\mu\right| = 1$ with
$\mbox{Im}(\mu)\neq 0$ at $\sigma = \sigma^{NS}$. The relation between
$\sigma^{NS}$ and the pair $\{\sigma_1^c,\sigma_2^c\}$ of critical
values depicted in~\fref{fig:histereses}(a) will be clarified below.

Like the AH bifurcation, the NS bifurcation also comes in two
different flavours: in the supercritical case, a stable closed
invariant curve (CIC --- the discrete-time analog of a limit cycle) is
born at $\sigma^{NS}$ and grows continually from zero amplitude; in
the subcritical case, an unstable CIC exists below $\sigma^{NS}$ and
engulfs the fixed point $\vec{P}^*$ at $\sigma^{NS}$ (above which the
system is typically attracted to another pre-existing, but stable,
CIC). Since the order parameter $q$ increases with the amplitude of
the oscillations (which is, roughly speaking, proportional to
$||\vec{\eta} ||$), a supercritical (subcritical) NS bifurcation in
the mean-field equations is suggestive of a continuous (discontinuous)
phase transition in the system (see e.g.~\cite{Assis09}).

The sign of the first Lyapunov coefficient $l_1$~\cite{Kuznetsov98}
indicates if the Neimark-Sacker bifurcation is supercritical~($l_1 <
0$) or subcritical~($l_1 > 0$). Its calculation is briefly reviewed in
the Appendix. We now have the necessary tools to unveil the complete
phase diagram.

% \begin{eqnarray}
% l_1 &=& \frac{1}{2}{\textrm{Re}}\left\{e^{-i\theta{}_0} \left[ \left< p, C(q,q,\bar{q}) \right> + 2\left< p, B(q,(I-A)^{-1}B(q,\bar{q})) \right>\right.\right. \nonumber\\
%      && + \left.\left.\left< p, B(\bar{q},(e^{2i\theta{}}I - A)^{-1}B(q,q)) \right>\right]\right\}
% \label{eq:firstLyap}
% \end{eqnarray}

\subsection{Phase diagram}

\begin{figure}[ht!]
\begin{center}
    \includegraphics[width=0.8\columnwidth]{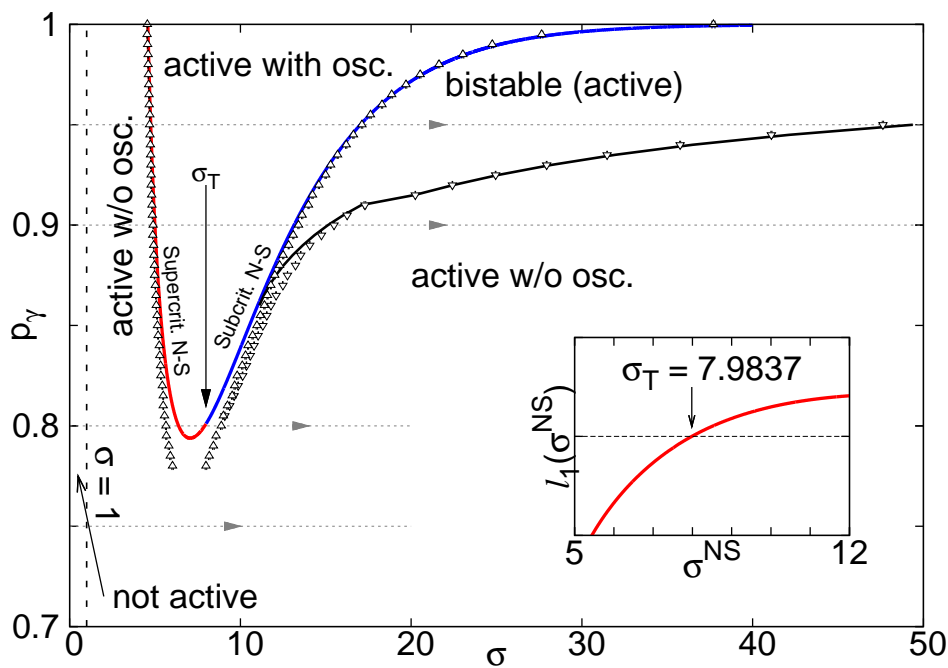}
    \caption{\label{fig:diagrama} Phase diagram for the complete graph
      (triangles) with $N=10^6$ (mean over $50$ runs, $t_{max}=10^3$,
      $t_{trans}=5\times{}10^2$). Solid red (blue) line is the
      supercritical (subcritical) Neimark-Sacker bifurcation predicted
      from linear analysis. Black line marks the (discontinuous)
      stability limit of the oscillating phase (as predicted by mean
      field). Note that the black and blue lines approach each other
      very closely before merging at $\sigma_T$.  Inset: first
      Lyapunov coefficient $l_1(\sigma{})$.}
\end{center}
\end{figure}

We have run simulations of complete graphs with $N=10^6$ excitable
units and employed the protocol described in
section~\ref{results.cg.sim} with $t_{max}=10^3$ to detect the width
of the hysteresis loop (coexistence region). We have tested and
verified that longer values of $t_{max}$ do not change our results
significantly. The phase diagram thus obtained from the simulations is
shown with symbols in \fref{fig:diagrama} (the horizontal gray lines
show the values of $p_\gamma$ used in \fref{fig:histereses}). To
obtain the NS lines of the mean-field equations, we have numerically
explored a special test function~\cite{Kuznetsov98}
$\Phi_{NS}(\sigma;p_\gamma) = \prod_{m < n}^{\tau} (1 - \mu_n\mu_m)$,
which changes its sign at the NS bifurcation.  As the first Lyapunov
coefficient $l_1$ determines whether the bifurcation is super- or
subcritical, its value along the bifurcation line is shown in the
inset of~\fref{fig:diagrama} (changing sign at $\sigma_T$). The solid
lines in~\fref{fig:diagrama} show the supercritical (red) and
subcritical (blue) bifurcation curves where
$\Phi_{NS}(\sigma^{NS};p_\gamma) = 0$.

Comparing the solid lines with the symbols in
\fref{fig:diagrama}, we observe that linear stability analysis
accurately accounts for the transition {\em from\/} an active (but
non-oscillating) phase {\em to\/} an oscillating phase. In other
words, it correctly predicts the lines $\sigma_c(p_\gamma)$ and
$\sigma_1^c(p_\gamma)$, where in both cases the non-oscillating phase
loses stability.

The transition at $\sigma_2^c(p_\gamma)$, however, cannot be predicted
by linear analysis. Note that in this case it is the stable CIC that
loses its stability (that of the fixed point $\vec{P}^*$ remaining
intact). This hints at the existence of a global bifurcation, which
can be numerically detected in the MF equations by direct iteration of
the map determined
by~equations~\eref{eq:MFDeterm},~\eref{eq:MFProbab}~and~\eref{eq:MFActive}.
To compare the $\tau$-dimensional MF map with system simulations, we
rewrite the complex vector $Z$
from~\eref{eq:complexKuramoto}~and~\eref{eq:norm_cond} as
\begin{equation}
  Z(t) = 1 + \sum_{s=1}^{\tau}P_t(s)(e^{i\phi_s}-1) \;,
\end{equation}
where $\phi_s = 2\pi{}s/({\tau+1})$. We can thus calculate $q$ for the
MF map and subject it to the same protocol used for detecting the
coexistence region in the simulations. The black solid line in
\fref{fig:diagrama} shows the $\sigma_2^c(p_\gamma)$ obtained by
iteration of the map, which is in good agreement with simulations
(symbols).

Note that in the lower part of the oscillating phase ($p_\gamma
\lesssim 0.8$) the order parameter detects oscillations in the
simulations which are not predicted by the mean-field analysis. This
phenomenon is due to {\em stochastic oscillations\/}, as recently
explained by Risau-Gusman and Abramson~\cite{Risau-Gusman07}: the
fixed point in the conflicting region is in fact {\em stable\/}, but
with an eigenvalue with a nonzero imaginary part. Inevitable
fluctuations throw the system away from the stable point, to which it
returns in spiral-like trajectories, yielding a nonzero $q$ even for
very large system sizes~\cite{Risau-Gusman07,Assis09}. 

For $p_\gamma=1$, we recover a quenched variant of the model by Girvan
et al.~\cite{Girvan02}. In this regime where intrinsic transitions are
deterministic, increasing the coupling will only reinforce collective
oscillations, and the fixed point $P_1^*$ never regains stability
(i.e. $\sigma_1^c\to\infty$). This suggests that even small amounts of
noise in the intrinsic dynamics of excitable elements can lead to
qualitatively different collective behavior in a regime of strong
coupling.

\section{\label{results.rg}Random graph}

\subsection{Mean-field and simulation results}

To understand network topology effects on synchronization, we study a
bidirectional random graph similar to an
Erd{\H{o}}s-R{\'{e}}nyi~\cite{Erdos} network, where $NK/2$ links
connect randomly chosen pairs~\cite{Kinouchi06a} and remain frozen
(``quenched'') throughout each run (and in each run, a new realization of
the network is created). The main difference with respect to the
complete graph (CG) is that in the random graph (RG) the value of
$\sigma$ is bounded from above: $\sigma\leq K$
[see~\eref{eq:pinf}]. The mean-field calculations, however, are
otherwise similar to that of the complete graph, with~\eref{eq:pinf}
replacing~\eref{eq:pinf_limit}. We therefore applied to the RG problem
the same procedures for determining the stability of the solutions,
the nature of the NS bifurcation and the boundary of the bistability
region (see section~\ref{results.cg}).

\begin{figure}[!htb]
\begin{center}
    \includegraphics[width=0.6\textwidth]{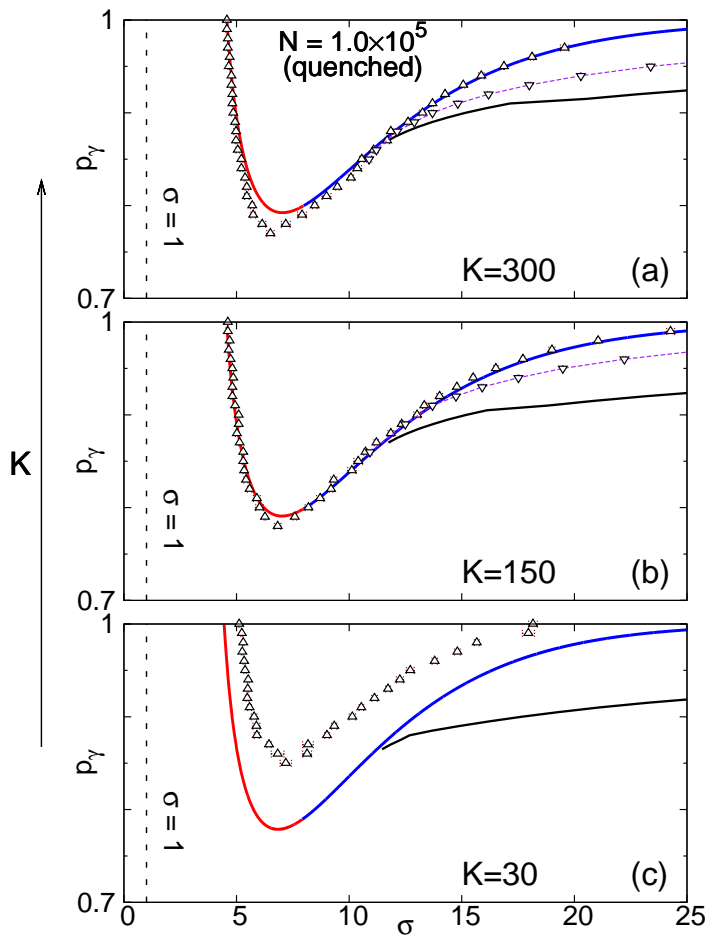}
    \caption{\label{fig:diagramaRGquenched} Phase diagram for
      (quenched) random graphs with (a) $K=300$, (b) $K=150$ and (c)
      $K=30$. Solid red (blue) line is the supercritical (subcritical)
      Neimark-Sacker bifurcation predicted from linear analysis. Black
      line marks the (discontinuous) stability limit of the
      oscillating phase (as predicted by mean field). Symbols are
      obtained from simulations (mean over $5$ runs) with
      $t_{max}=3\times 10^3$ steps ($t_{trans}=2\times 10^3$ steps)
      and $N=10^5$. Standard errors are smaller than symbol size. The
      purple dashed line is a guide for the eyes and marks the
      stability limit of the oscillating phase for simulations.}
\end{center}
\end{figure}

Given their uncorrelated assigment of links and short distances among
sites, random graphs are usually regarded as the natural topology in
which mean-field predictions are expected to hold. Indeed, simulations
and mean-field calculations agree nearly perfectly as far as the phase
transition at $\sigma=1$~\cite{Kinouchi06a} is
concerned. In~\fref{fig:diagramaRGquenched}(a)-(b) we see that a good
agreement is also observed in the transitions to a synchronized phase
for large values of $K$. Note, however, that simulations and
mean-field predictions differ at the rightmost boundary of the
bistable region, and the disagreement worsens as $K$ decreases. As
shown in~\fref{fig:diagramaRGquenched}(c), for smaller values of $K$
bistability was not even detected in the simulations, and the
oscillating phase is substantially smaller than predicted by mean
field.

\subsection{Annealed random graphs}

\begin{figure}[!htb]
\begin{center}
    \includegraphics[width=0.6\textwidth]{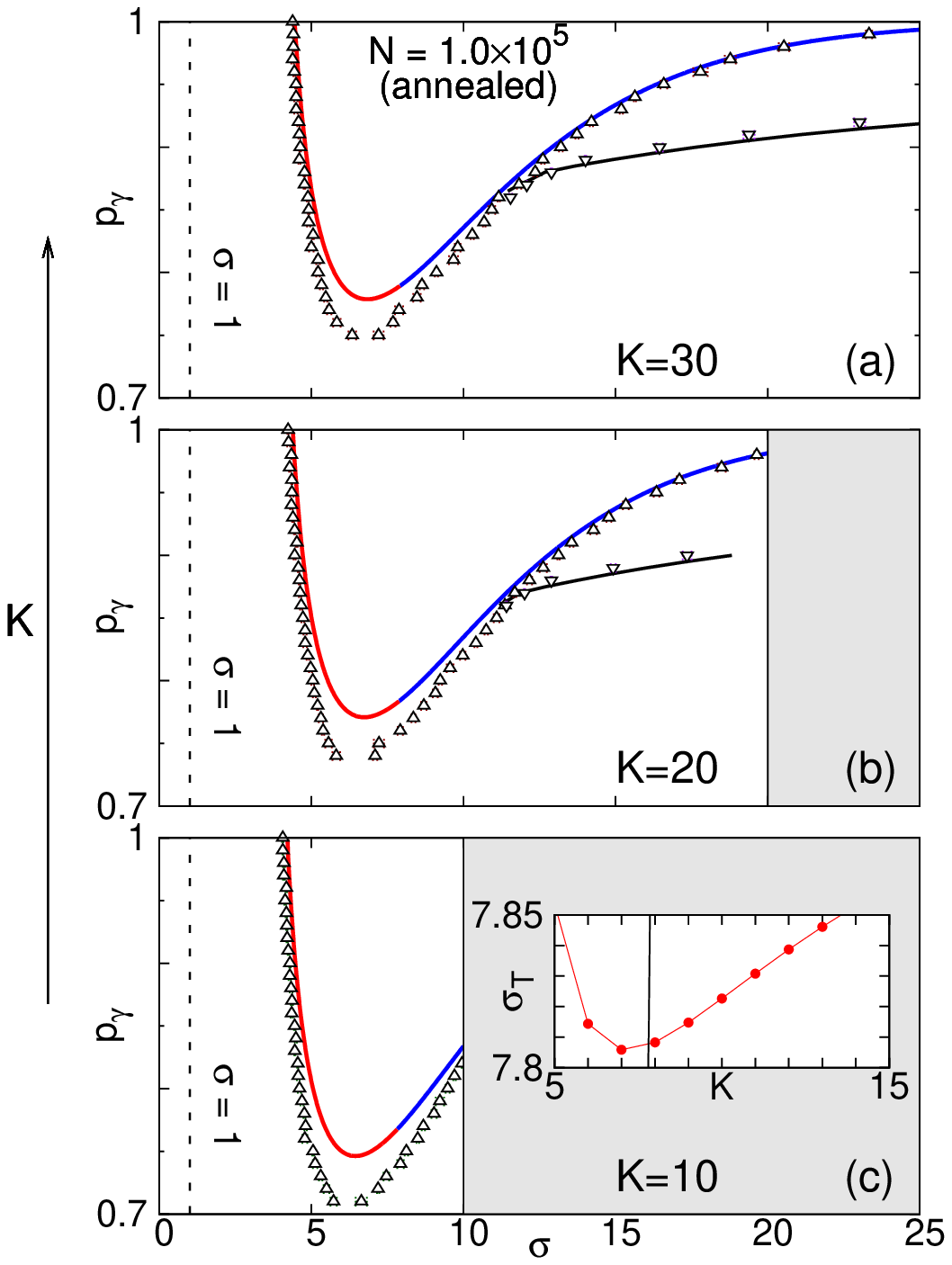}
    \caption{\label{fig:diagramaRGannealed} Phase diagram for
      (annealed) random graphs with (a) $K=30$, (b) $K=20$ and (c)
      $K=10$. Solid red (blue) line is the supercritical (subcritical)
      Neimark-Sacker bifurcation predicted from linear analysis. Black
      line marks the (discontinuous) stability limit of the
      oscillating phase. Symbols are obtained from simulations (mean
      over $5$ runs) with $t_{max}=3\times 10^3$ steps ($t_{trans}=2\times{}10^3$ steps) and
      $N=10^5$. Standard errors are smaller than symbol size. Grey
      shaded areas correspond to a forbidden region where
      $\sigma>K$. Inset: $\sigma_T$ for different values of $K$,
      showing the existence of a $K_c = \sigma_T(K_c) = 7.8074(1)$
      below which no bistable region exists.}
\end{center}
\end{figure}

Could correlations (which are neglected by the mean-field
approximation) account for the discrepancy observed
in~\fref{fig:diagramaRGquenched}? In order to assess the role of the
correlations associated with the quenched connectivity, we studied an
annealed variant of the model where the $K$ neighbors of each site are
randomly chosen at each time step~\cite{Girvan02,Sinha07}. Results are
shown in~\fref{fig:diagramaRGannealed}, in which we restrict
ourselves to smaller values of $K$ because results are essentially
indistinguishable from the quenched case for large $K$.

For smaller $K$, three features are noteworthy.  First, the agreement
with mean field results is recovered (apart from the stochastic
oscillations in the lower end of the oscillating phase, like in the
previous cases). This therefore confirms the suspicion that
correlations associated to the quenched connectivity can indeed
undermine collective oscillations. This is not surprising, given the
difficulty of establishing collective oscillations of excitable
elements in hypercubic lattices~\cite{Kuperman01,Assis09}.

Second, the bound $\sigma<K$ impoverishes the repertoire of detected
phenomena for small $K$ [see the forbidden gray regions
in~\fref{fig:diagramaRGannealed}(b) and (c)]. Note that the
coexistence region shrinks as $K$ decreases. In fact, since $\sigma_T$
varies very slowly with $K$ [see the inset
of~\fref{fig:diagramaRGannealed}(c)], there is a minimum value of $K$
[satisfying $K_c = \sigma_T(K_c)$] below which the system shows no
bistability (i.e. there is no change in the sign of $l_1$). We have
numerically estimated $K_c=7.8074(1)$.

Finally, note that the oscillating phase extends into lower values of
$p_\gamma$ for {\em decreasing\/} $K$. This is a rather
counter-intuitive result. It means that, for fixed $\sigma$ and
$p_\gamma$, it is possible to take the system from a non-oscillating
to an oscillating phase by lowering the connectivity $K$. This is a
particularity of the annealed RG (and the mean-field approximation),
however. Note in~\fref{fig:diagramaRGquenched} that the opposite (and
expected) trend is observed for quenched random graphs. It remains to
be studied whether refining the approximation (including
e.g. first-neighbor correlations~\cite{Furtado06,Rozhnova09c}) can
reconcile mean-field with quenched random graph results.

\subsection{\label{fsizescaling}Finite-size scaling}

\begin{figure}[ht]
\begin{center}
\includegraphics[width=0.8\textwidth]{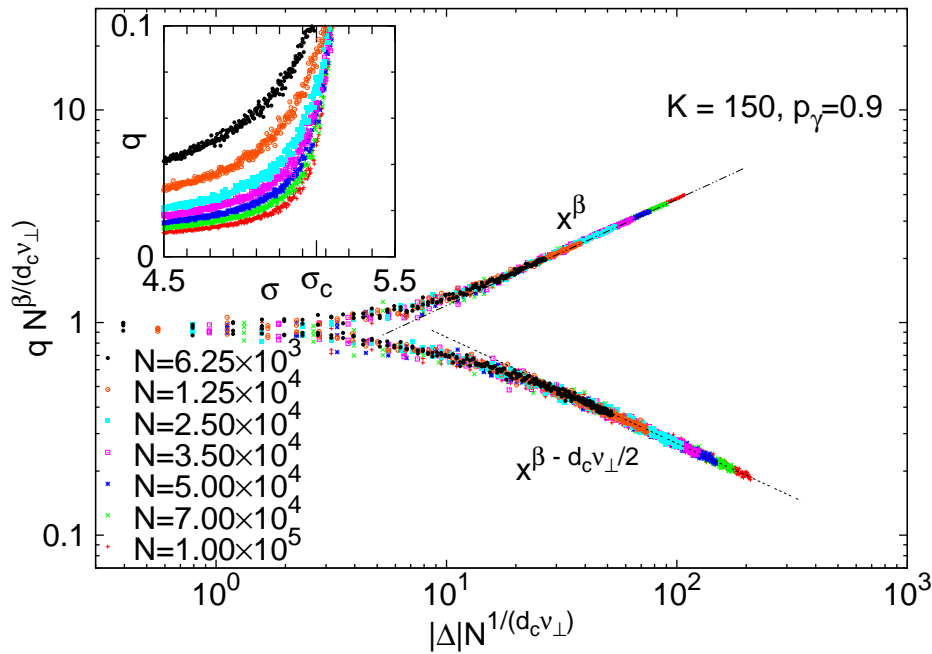}
\caption{\label{fig:collapseProb} Collapse in a probabilistic
  ($p_\gamma = 0.9$) regime for $K=150$, $\sigma_c = 5.16$. We used
  $\beta = 1/2$, $\nu_\perp = 1/2$, $d_c = 4$.  Mean over $15$ runs,
  with standard errors smaller than symbol size. Other parameters were
  $t_{max}=10^4$,$t_{trans}=7\times{}10^3$,
  $\delta\sigma=5\times{}10^{-3}$}
\end{center}
\end{figure}

Our phase diagrams rely heavily on the proposed order parameter
$q$. The inset of \fref{fig:collapseProb} indicates that near
$\sigma_c$ its behavior becomes increasingly abrupt with increasing
system size $N$. In order to confirm that $q$ indeed possesses the
basic properties of a {\it bona fide\/} order parameter, here we show
that at the supercritical NS bifurcation (i.e. a second order phase
transition) it satisfies the scaling relations that would be expected
from standard finite-size scaling (FSS) theory.

Defining $\Delta\equiv \sigma-\sigma_c$, FSS predicts that $q \propto
L^{-\beta/\nu_\perp}f\left(\Delta L^{1/\nu_\perp}\right)$ for a
lattice with linear size $L$~\cite{Marro99}, where the critical
exponents are defined as $q \propto |\Delta|^\beta$ and $\xi \propto
|\Delta|^{-\nu_\perp}$ in the limit $N=L^d \to\infty$ (where $\xi$ is the
correlation length). This holds for $d$ below the upper critical
dimension $d_c$. For $d\geq d_c$, mean-field exponents are expected
and the scaling relation has to be modified~\cite{Brezin82}
with $L\to N^{1/d_c}$, so
\begin{equation}
\label{eq:fss}
q \propto N^{-\beta/d_c\nu_\perp} f\left(\Delta
  N^{1/d_c\nu_\perp}\right)\; .
\end{equation}
This modified version of the usual FSS relation was shown to also hold
for infinitely coordinated networks~\cite{Botet82} (i.e. the complete
graph).

For large argument, the scaling function in~\Eref{eq:fss} becomes
$f(x) \propto x^{\beta}$, as usual~\cite{Marro99}. In the subcritical
regime ($\Delta < 0$), one expects $q\sim
\mathcal{O}(N^{-1/2})$~\cite{Strogatz00}, so $f(\Delta
N^{1/d_c\nu_\perp}) \propto N^{-1/2} N^{\beta/d_c\nu_\perp}$. For this
to be true, $f(x) \propto x^{\beta - \frac{d_c\nu_\perp}{2}}$ when
$x<0$. \Fref{fig:collapseProb} shows an excellent data collapse for
(quenched) random graphs with different system sizes. Consistent power
laws are obtained with standard mean-field exponents (as expected for
random graphs), namely $\beta=1/2$~\cite{Strogatz00}, $\nu_\perp=1/2$
and $d_c=4$~\cite{Marro99}. Similar results are obtained with complete
graphs (not shown).

\section{\label{conclusions}Concluding remarks}

We have studied the effects of a probabilistic refractory period in
the collective behavior of a large number of coupled excitable
cellular automata. We have obtained the mean-field solution of the
model and compared it with simulations of complete as well as random
graphs. The continuous phase transition to a synchronized regime is
associated to a Neimark-Sacker bifurcation in the mean-field
equations. 

This scenario is similar to what has been previously obtained by
Girvan et al.~\cite{Girvan02} in a model of deterministic excitable
automata ($p_\gamma=1$ in our model). The effects of setting
$p_\gamma<1$, however, are drastic, and appear for sufficiently strong
coupling $\sigma$, when we have observed that oscillations
vanish. This is in contrast with the transition to the {\em
  absorbing\/} state found in Ref.~\cite{Girvan02} for strong
coupling. While in their model the transition is due to very large
amplitudes driving the system into rest ($P_t(1) = 0$), in our model
the system is thrown into an {\em active\/} albeit disordered state,
which still has $P_t(1) \neq 0$, but no oscillations.

Furthermore, only for non-deterministic excitable elements do we
observe bistability, with an oscillating and an active (but
non-oscillating) phase coexisting. This leads to hysteresis cycles,
whose sizes can depend on the system size (notably for the complete
graph) and eventually disappear in random graphs with small enough
mean connectivity $K$. 

Although we have restricted ourselves to $\tau=3$, preliminary results
suggest that the overall scenario is preserved for larger values of
$\tau$, specially regarding the first transition at $\sigma_c$. The
observation of bistability and hysteresis is more difficult for larger
values of $\tau$, owing to stronger finite-size effects. These are
similar to those reported by Girvan et al.: for finite $N$ and
sufficiently strong coupling, the CIC grows in amplitude and nears the
absorbing state, to which the system is thrown by
fluctuations~\cite{Girvan02}. Distinguishing between that type of
transition and the bistability reported here is not obvious and
remains to be studied.

It is interesting to note that a simple model allows the
straightforward application of standard techniques of nonlinear
dynamics to the study of these phase transitions, which could
otherwise be difficult to tackle. Note that in the limit $p_\gamma=
1$, each excitable unit in our model has, at the end of its refractory
period, a perfect memory of its past $\tau$ time steps. Incorporating
this memory in a continuous-time model would require non-Markovian
dynamics, as recently proposed by Gon{\c{c}}alves et
al.~\cite{Abramson10a}. Interestingly, their model also shows an
oscillating phase with reentrance, whose size decreases as memory time
decreases. In our model, this corresponds to lowering $p_\gamma$, with
a similar effect on the collective behavior. It remains to be
investigated whether bistability also occurs in non-Markovian
continuous-time models.

Finally, a note of caution is in order regarding the scaling results
of~\fref{fig:collapseProb}. The fact that a data collapse is obtained
employing an upper critical dimension $d_c=4$ by no means implies that
a lower critical dimension exists. In fact, we are not aware of models
which exhibit collective oscillations of excitable elements in
hypercubic lattices (though they do appear if stimulated with a
Poisson drive~\cite{lewis00}). It remains to be investigated whether
the results of this model hold for small-world networks and other
complex topologies~\cite{Kuperman01,Gade05}, which are extremely
appealing for applications in Neuroscience. This is currently under
investigations (results will be published elsewhere).

In summary, we have shown that collective oscillations of excitable
elements have robustness to a certain degree of stochasticity in their
intrinsic dynamics. For fixed coupling, there is a critical value of
$p_\gamma$, below which no oscillations are stable. Fixing $p_\gamma
<1$ and increasing the coupling $\sigma$, on the other hand, leads to
interesting new phenomena such as bistability and discontinuous
transitions. Taken together, our results suggest that even weakly
noisy dynamics can qualitatively change the collective oscillating
behavior. Studies attempting to verify whether these phenomena are
observed in more detailed excitable networks (e.g. modelled by
stochastic differential equations) would certainly be welcome.

\ack

The authors gratefully acknowledge enlightening discussions with
Vladimir R. V. Assis, as well as finantial support from Brazilian
agencies CNPq, FACEPE, CAPES and special programs PRONEX (F{\'\i}sica
Biol\'ogica) and INCeMaq.

\appendix

\section[\hspace{2.3cm}Lyapunov coefficient]{Lyapunov coefficient}

%
% Kuznetsov, 98 - pg. 186 - eq. 5.74
%

Let $\vec{u}$ and $\vec{v}$ be respectively the right and left (adjoint)
eigenvector of the Jacobian matrix:
\begin{eqnarray}
A\vec{u} &=& e^{i\theta{}_0}\vec{u} \;, \\
A^T\vec{v} &=& e^{-i\theta{}_0}\vec{v} \;,
\end{eqnarray}
with both normalized: $\left<\vec{v},\vec{u}\right> =
1=\left<\vec{u},\vec{u}\right>$ (brackets denote the standard complex
inner product). Let also $\vec{B}(\vec{x},\vec{y})$ and
$\vec{C}(\vec{x},\vec{y},\vec{z})$ be multilinear functions
proportional to the first nonlinear terms of the Taylor expansion of
$\vec{F}$ at $\sigma=\sigma^{NS}$, i.e.,
\begin{equation}
B_j(\vec{x},\vec{y}) = \left. 
  \sum_{k,l=1}^{\tau}{\frac{\partial^2F_j(\vec{\xi};\sigma^{NS})}{\partial\xi_k\partial\xi_l}} 
\right|_{\vec{\xi} = \vec{P}^{*}} \!\!\!\!\!\! x_ky_l \;,
\end{equation}
\begin{equation}
C_j(\vec{x},\vec{y},\vec{z}) = \left. 
  \sum_{k,l,m=1}^{\tau}{\frac{\partial^3F_j(\vec{\xi};\sigma^{NS})}{\partial\xi_k\partial\xi_l\partial\xi_m}} 
\right|_{\vec{\xi} = \vec{P}^{*}} \!\!\!\!\!\! x_ky_lz_m \;.
\end{equation}
If we now define
\begin{eqnarray}
\vec{r} &=& (I - A)^{-1}\vec{B}(\vec{u},\bar{\vec{u}}) \;, \\
\vec{s} &=& (e^{2i\theta{}}I - A)^{-1}\vec{B}(\vec{u},\vec{u}) \;,
\end{eqnarray}
where $I$ is the $\tau{}\times{}\tau{}$ identity matrix and
$\bar{\vec{u}}$ is the conjugate of $\vec{u}$, the
coefficient $l_1$ is finally given by~\cite{Kuznetsov98}
\begin{equation}
l_1 = \frac{1}{2} {\textrm{Re}} \left\{
  e^{-i\theta{}_0} \left[ 
    \left< \vec{v}, \vec{C}(\vec{u},\vec{u},\bar{\vec{u}}) \right> 
    + 2\left< \vec{v}, \vec{B}(\vec{u},\vec{r}) \right> 
    + \left< \vec{v}, \vec{B}(\bar{\vec{u}},\vec{s}) \right>
  \right]
\right\} \;.
\label{eq:firstLyap}
\end{equation}

\section*{References}
\bibliography{copelli,sync}
\end{document}